\documentclass[fleqn,usenatbib]{mnras}

\usepackage{newtxtext,newtxmath}

\usepackage[T1]{fontenc}

\usepackage{graphicx}	
\usepackage{amsmath}	
\usepackage{url}

\title[FRBs as scaled-up solar radio bursts]{Galactic and cosmological fast radio bursts as scaled-up solar radio bursts}

\author[Wang, Zhang and Dai]{
    F. Y. Wang,$^{1,2}$\thanks{E-mail: fayinwang@nju.edu.cn}
    G. Q. Zhang,$^{1}$
    Z. G. Dai$^{1,2}$
    \\
    $^{1}$School of Astronomy and Space Science, Nanjing University, Nanjing 210093, China\\
    $^{2}$Key Laboratory of Modern Astronomy and Astrophysics (Nanjing University), Ministry of Education, Nanjing 210093, China
}

\date{Accepted XXX. Received YYY; in original form ZZZ}

\begin{document}
    \label{firstpage}
    \pagerange{\pageref{firstpage}--\pageref{lastpage}}
    \maketitle

\begin{abstract}
Fast radio bursts (FRBs) are bright milliseconds radio transients
with large dispersion measures. Recently, FRB 200428 was detected in temporal coincidence with a hard X-ray flare from the
Galactic magnetar SGR 1935+2154, which supports that at least some FRBs are from magnetar activity.
Interestingly, a portion of X-ray flares from magnetar XTE J1810-197 and the Sun are also accompanied
by radio bursts. Many features of Galactic FRB 200428 and cosmological FRBs resemble solar radio bursts.
However, a common physical origin  among FRBs, magnetar radio pulses and solar radio bursts has not yet
been established. Here we report a universal correlation between X-ray luminosity
and radio luminosity over twenty orders of magnitude among solar type III radio bursts, XTE J1810-197 and Galactic FRB 200428. This universal correlation reveals that the energetic electrons which produce the
X-ray flares and those which cause radio emissions have a common origin, which can give stringent limits on the generation process of radio bursts. Moreover, we find similar occurrence frequency distributions of energy, duration and waiting time for solar radio bursts, SGR 1935+2154 and repeating FRB 121102, which also support the tight correlation and the X-ray flares temporally associated with radio bursts. All of these distributions can be understood by avalanche models of self-organized criticality systems. The universal correlation and statistical similarities indicate that the Galactic FRB 200428 and FRBs seen at cosmological distances can be treated as scaled-up solar radio bursts.
\end{abstract}


\section{Introduction}
Fast radio bursts (FRBs) are intense radio transients with extreme high
brightness temperatures that show dispersion relations to be consistent
with propagation through cold plasma \citep{Lorimer2007,Thornton2013}.
Until now, more than one hundred FRBs have
been discovered. All previous localized
FRBs were at cosmological distances \citep{Chatterjee2017,Bannister2019,Prochaska2019,Ravi2019,Marcote2020,Macquart2020}.
Even with precise localizations
in their host galaxies, the physical origin of FRBs is still a
mystery \citep{Platts2019,Zhang2020}. Magnetars have been proposed as the possible engine to
power repeating bursts from FRB sources \citep{Popov2013,Kulkarni2014,Murase2016,Katz2016,Metzger2017,Wang2017,Beloborodov2017,Yang2018,Wang2020}.

The situation is dramatically changed with the
recent discovery of FRB 200428
from the Galactic soft-gamma repeater (SGR) 1935+2154 independently by the Canadian Hydrogen Intensity Mapping
Experiment \citep{CHIME/FRB2020} and the Survey
for Transient Astronomical Radio Emission 2 (STARE2) \citep{Bochenek2020}.
Meanwhile, a hard X-ray burst from SGR 1935+2154 was temporally coincident with this FRB, which has been observed
by many telescopes \citep{Li2020,Tavani2020,Ridnaia2020,Mereghetti2020}.
The fluence of this FRB at 1.4 GHz is  1.5$\pm 0.3$ Mega-Jy ms \citep{Bochenek2020}. This burst
is about a factor of  40 less energetic than the weakest
burst detected from localized cosmological
FRB sources, but a factor of $4\times 10^3$ more energetic than any millisecond radio burst previously
observed in the Milky Way. This discovery strongly supports the models based on magnetars that have been proposed
for cosmological FRBs, and also proves the physical connection between X-ray bursts and radio bursts.
Only a small fraction of X-ray flares is accompanied by luminous FRBs, based on the fact that
29 of the X-ray flares from SGR 1935+2154 have no radio signal detected down to mJy ms by Five-hundred-meter Aperture Spherical radio Telescope \citep{Lin2020}.
Recently, bright X-ray pulses in temporal coincidence with radio pulses were detected from Galactic magnetar XTE J1810-197 \citep{Pearlman2020}.
Although the radio pulses have lower energies compared with FRBs, they have similar characteristics to radio bursts detected from FRBs, such as millisecond duration and frequency structure. This suggests active magnetars can generate radio and X-ray emissions in a large energy range.

On the other hand, it is well known that solar type III radio bursts are usually associated with solar X-ray flares \citep{Bastian1998}.
Solar type III radio bursts, identified by high
brightness temperatures and rapid frequency drift, are a common
signature of fast electron beams generated during
solar flares. They are produced by coherent emissions arising from the
nonlinear conversion of Langmuir waves generated by two-stream
instability of electron beams \citep{Ginzburg1958,Melrose2017}.
Observations show that about 30\% hard X-ray flares are temporally correlated with type III radio bursts \citep{Reid2017}.
There are at least four common properties
for FRBs and solar type III radio bursts. First, they both
have high brightness temperatures, 10$^6$ K-10$^{12}$ K for solar
radio bursts and as high as 10$^{35}$ K for FRBs. Second, the
frequency drift (high-to-low temporal evolution) is found in
repeating FRBs \citep{Hessels2019}, and solar type III radio
bursts. Third, both radio bursts show similar
intensity temporal
evolution \citep{Bastian1998,Hessels2019}. Fourth, both radio bursts
have been found to be temporally correlated with X-ray flares. \citet{Lyutikov2002}
already proposed that magnetar radio/X-ray flares are similar to solar flares, triggered by the magnetic
field instabilities in the magnetars' magnetospheres. In this model, because the radio
emission is generated by the electrons accelerated at the reconnection
site, the intensity and profile of radio bursts will
be strongly correlated with the X-ray bursts, which is confirmed by the two peaks of FRB 200428 and the associated
X-ray flare peaks.

Although radio bursts are common phenomena in FRB sources, magnetars and the
Sun, the burst luminosity spans more than 20 orders of magnitude. They also show similar features. However, a physical
analogy among FRBs, magnetar radio pulses and solar type III radio bursts has not yet
been established. It is generally accepted that the energetic electrons which produce the
hard X-ray flares and those which cause the type III
radio emissions during a solar flare have a common origin, which indicates that the X-ray flare emissions are correlated with
the radio burst emissions \citep{Bastian1998}. Whether this correlation exsits in magentars and FRB sources is unclear.

Power-law size distributions are found in many astrophysical systems \citep[for a review, see][]{Aschwanden2016}. Solar X-ray flares have been found to show power-law size distributions \citep{Crosby1993}. Giant pulses of several pulsars show power-law size distributions \citep{Cognard1996,Lundgren1995,Cairns2004}. The size distributions of the fluences of SGR 1900+14 and SGR 1806-20 are
found to exhibit power-law distributions with slopes around $\alpha_E\sim 1.6$ \citep{Gogus1999,Gogus2000}. So it urgent to know whether FRBs and the associated X-ray bursts of magnetars show similar size distributions.

In this paper, we will show that the three kinds of radio bursts and the accompanied X-ray flares follow a universal correlation between X-ray luminosity $L_X$ and radio luminosity $L_R$. Moreover, the radio bursts and the accompanied X-ray flares of FRB sources and the Sun show similar energy, duration and waiting time distributions. The structure of this paper is arranged as follows. In next section, the data sample are given. In section 3, we show the results. Summary is given in section 5.

\section{Data sample}

\subsection{Data used in $L_X-L_R$ correlation}
The detection by CHIME in 400-800 MHz band reveals
that FRB 200428 has two pulses, each about 0.5 ms wide and separated by 28.91 ms \citep{CHIME/FRB2020}. So the duration is about 30 ms. The peak flux densities for the two peaks are 110 kJy and 150 kJy \citep{CHIME/FRB2020}, respectively. The observation by STARE2 shows  that the peak fluence in the 1281-1468 MHz is 1.5$\pm$0.3 MJy ms \citep{Bochenek2020}. The flux difference may be duo to the complex spectral feature. Because the two bands are very close, we adopt the peak fluence from STARE2 and the duration from CHIME. The peak radio luminosity is 2.61$\times 10^{27}$ erg s$^{-1}$ Hz$^{-1}$. The associated X-ray flare has a peak flux of $~ 6\times 10^{-6}$ erg cm$^{-2}$ s$^{-1}$ in 20-200 keV band \citep{Mereghetti2020}. The distance to SGR 1935+2154 can range from 6.6 kpc to 12.5 kpc \citep{Kothes2018,Zhou2020,Zhong2020}. In this paper, $6.6\pm 0.7$ kpc is used \citep{Zhou2020}. The peak luminosity for the X-ray flare is 3.13$\times 10^{40}$ erg s$^{-1}$.

We also used the fluxes of type III radio bursts and X-ray flares given by Reid \& Vilmer (2017) \citep{Reid2017}. The 164 MHz flux for radio bursts and 25-50 keV flux for X-ray flares are selected, where the peaks in the two bands must be within 40 s of each other. Because the radio luminosity at 1.4 GHz of FRB 200428 is used, the 164 MHz flux is scaled to 1.4 GHz by assuming a spectral index of -1.78 \citep{Reid2017}. When calculating the X-ray flux, the spectrum $dN/dE \propto E^{-3.9}$ at 25-50 keV is used \citep{Benz2010}.

XTE J1810-197 was discovered by the Rossi X-ray Timing Explorer in 2003 \citep{Ibrahim2004}. The distance to this magnetar is about 4 kpc \citep{Minter2008}. Recently, simultaneous radio pulses by NASA DSN 34m radio telescopes and X-ray pulses by Neutron star Interior Composition Explorer (NICER) are detected from XTE J1810-197 \citep{Pearlman2020}. The observational bands are 4.8 GHz and 0.5-10 keV for radio and X-ray, respectively. We use the parameters of radio and X-ray bursts shown in figure 2 of Pearlman et al. (2020).  The number of bursts is 19. The area of NICER X-ray Timing Instrument is 1900 cm$^2$. The spectral index is from -0.5 to 0 between 1.4 and 144 GHz \citep{Pearlman2020}. In calculation, we assume a flat spectrum to obtain the radio flux density at 1.4 GHz. The radio pulses from XTE J1810–197 share similarities with the radio
bursts from repeating FRB sources, such as millisecond duration, frequency structure and narrow emission range. It must be noted that the X-ray pulses from XTE J1810–197 are temporally distinct and less energetic than giant flares and X-ray bursts previously observed from magnetars \citep{Pearlman2020}, indicating that they display highly
variable X-ray emission. But the processes responsible for this behavior
remain a mystery, which encourages us to build a unified picture. The X-ray band is different from that of SGR 1935+2154 flare. According to the X-ray burst of XTE J1810–197 observed in 2018,
the luminosity in 2-10 keV band is comparable to that in the high-energy band \citep{Gotthelf2019}. Therefore, we assume that the luminosity in 2-10 keV band is similar as the high-energy one.

\subsection{Data used in frequency distributions}

The FRB 121102 has the largest burst sample. We collect the bursts of FRB 121102 from the observation by Green
Bank telescope at 4-8 GHz. Recent work identified 93 pulses of FRB
121102 from 6 hours of observation
\citep{Gajjar2018,ZhangY2018}. This
observation constructs the largest sample of FRB 121102 for a single
observation. Using this sample, we can avoid the complex selection
effect caused by the different telescopes at different frequencies.

For SGR 1935+2154, the bursts observed by Gamma-ray Burst Monitor on the Fermi Gamma-ray Space Telescope
from 2014 to 2016 are used \citep{Lin2020}. The number of bursts is 112.

As for solar type III radio bursts, we select the data from the
National Centers for Environmental Information (NCEI) observed by United States Air Force Radio Solar Telescope Network (RSTN)
\footnote{ftp://ftp.ngdc.noaa.gov/STP/space-weather/solar-data/solar-features/solar-radio/radio-bursts/reports/fixed-frequency-listings/},
which has observed for many years and has
accumulated lots of data. RSTN consists of four sites: Learmonth,
Palehua, Sagamore Hill and San Vito. The device and analysis methods
in all sites are identical, so we can simply put them together to
study their statistical properties. The data of RSTN contains solar
radio bursts at 8 frequencies (245 MHz, 410 MHz, 610 MHz, 1415 MHz,
2695 MHz, 4995 MHz, 8800 MHz, 15400 MHz). We divide the data into
multiple subsamples based on frequency and calculate their
statistical nature on each frequency. In order to obtain high
quality data, we filter the data according to the criteria of \citet{Giersch2017}.
Based on these criteria, we select a large sample of solar type
III radio bursts. The data of RSTN consists of
eight frequencies. In these frequencies, 4995 MHz is more
interesting, because it has the similar frequency with the data of
FRB 121102. We also found that the best-fitting parameters ($\alpha_E, \alpha_T, \lambda_0$) are consistent with
each other for the eight frequencies.

\section{Results}
\subsection{A universal correlation between X-ray and radio luminosities}
Figure 1 shows the correlation between peak X-ray luminosity ($L_X$ in units of ergs per second) and simultaneously peak radio luminosity ($L_R$ in units of ergs per second per hertz) for solar type III radio bursts, magnetar XTE J1810-197 \citep{Pearlman2020} and FRB 200428 \citep{Bochenek2020,Mereghetti2020}. Fitting the solar and XTE J1810-197 data, a tight correlation
\begin{equation}
\log L_X=(1.10\pm 0.01)\times \log L_R+(10.17\pm 0.10)
\end{equation}
with a linear regression coefficient R-square=0.993 is found. The data of FRB 200428 (red square) is dramatically consistent with this correlation. The similarity of the
luminosity ratio $L_X/L_R$, over twenty orders of magnitude in luminosity is
surprising. The correlation suggests that the energetic electrons which produce the
hard X-ray flares and those which generate
radio emissions have a common origin. This $L_X-L_R$ correlation
can be used to test the nature of a phenomenon and the mechanism of the associated radio emission. Some theoretical models also predict larger X-ray flares accompanying brighter FRBs \citep{Metzger2019,Lyutikov2020,Lu2020}. This correlation can be tested with future simultaneous observations of FRBs and X-ray flares from magnetars.
The dashed line in figure 1 shows the G\"{u}del-Benz correlation between $L_X$ and $L_R$ \citep{Guedel1993}, which only holds for gyrosynchrotron radio emissions and thermal X-ray emissions. For solar type III radio bursts, magnetar radio pulses and FRBs, the radio emissions are coherent emission, not gyrosynchrotron radiation. Therefore, they deviate from the G\"{u}del-Benz correlation.

\subsection{Occurrence frequency distributions}
Below, we discuss the frequency distributions of solar type III radio bursts, SGR 1935+2154 and FRB 121102, inspired by
the fact that solar radio bursts and FRB 200428 are both associated with X-ray flares. The tight correlation and X-ray flares temporally
associated with radio bursts also require the occurrence frequency distributions of energy, duration and waiting time for them are similar.

Figure \ref{fig:frbE} shows the energy frequency distributions of solar type III radio bursts, SGR 1935+2154 and FRB 121102.
For solar radio bursts, the differential distribution is used.
The number of bursts $N(E)dE$ with energy
between $E$ and $E + dE$ can be expressed by
\begin{equation}\label{energydis}
dN/dE\propto E^{-\alpha_E},
\end{equation}
where $\alpha_E$ is the power-law index. The numbers of bursts of FRB 121102 and SGR 1935+2154 are small, so it's preferable to
consider cumulative distribution \citep{Aschwanden2015}
\begin{equation}
\label{eq:Fculdis}
N(>E) = 1 + (N_{ev} - 1) \left( \frac{(E_2 + E_0)^{1 - \alpha_E} - (E + E_0) ^ {1 - \alpha_E}}{(E_2 + E_0)^{1 - \alpha_E} - (E_1 + E_0) ^ {1 - \alpha_E}} \right)
\end{equation}
rather than the differential
distribution. In the above equation, $ N_{ev} $ is the number of bursts, $ E_0 $ is a threshold value, $ \alpha $ is the power-law
index, $ E_1 $ and $ E_2 $ is the minimum energy and maximum energy, respectively. We use the uniform logarithmic bins to group these bursts
and the error of $i$th bin is $ \sqrt{N_i-N_{i+1}} $, where $ N_i $ and $N_{i+1}$ is the number of bursts in each bin \citep{Aschwanden2019}. This method takes only independent events into account for the estimation
of the uncertainty.
The Markov chain Monte Carlo (MCMC) technique is used to derive the best-fitting parameters. The best-fitting power-law indices are $\alpha_E= 1.51 \pm 0.19 $, $1.66\pm 0.06 $ and $1.82\pm0.20$ for solar radio bursts, FRB 121102 and SGR 1935+2154 with $1\sigma$ errors, respectively. The energy distributions for the three sources are consistent with each other at $1\sigma$ confidence level. The energy distribution of solar radio bursts is consistent with that of type III radio bursts observed by the Nancay Radioheliograph from 1998 to 2008 \citep{SaintHilaire2013}. The value of $\alpha_E$ for SGR 1935+2154 is consistent with other SGRs \citep{Cheng2020}.
However, a slight larger value of $\alpha_E=1.78\pm0.01$ is found by \citet{Lin2020} using an ideal power-law function. The reason is that the high-energy cutoff is considered in our fitting. \citet{Yang2020} also found the power-law index is $\alpha_E=1.72^{+0.13}_{-0.19}$ at high-energy range. The similar power-law indices between SGR 1935+2154 and FRB 121102 support that the tight correlation between radio burst luminosity and X-ray flare luminosity. We also try to use the broken power-law function to fit the energy distribution of FRB 121102. The
fitting result is shown as red dashed line in Figure \ref{fig:frbE}. The power-law index for the low energy band is $ \alpha_{E, low} = 1.75\pm0.64 $ and that of high energy end is $ \alpha_{E, high} = 2.38\pm0.51 $. The fit of broken power-law looks much better. But given the small number of bursts, the
error of the fitting results are large. The steepness of power-law slope may be associated
with finite-size effects in a self-organized criticality (SOC) system, which can cause an exponential-like cutoff at
the upper end of the size distribution, as it is found in many astrophysical systems \citep{Aschwanden2016}.

Figure 3 shows the differential distribution of duration $T$ for
solar type III radio bursts at 4995 MHz (left panel),  the
cumulative distributions of duration for FRB 121102 (middle panel) and SGR 1935+2154 (right panel),
respectively. Using the same fitting method as Figure 2, we find the
power-law indices are $\alpha_T= 1.69 \pm 0.02 $, $1.57
\pm 0.11 $ and $1.78\pm0.06$ for solar radio bursts, FRB 121102 and SGR 1935+2154, respectively.
The value of $\alpha_T$ for SGR 1935+2154 is consistent with those of other SGRs \citep{Wang2017}.
We can see that the duration distributions for the three sources are consistent with each other at $1\sigma$ confidence level.
The broken power-law function is also used to fit the duration distribution of SGR 1935+2154. We show the results in Figure \ref{fig:frbwidth} with
red dashed line. The index of low-energy band is $ \alpha_{T,low} = 1.62\pm0.54 $, which is consistent with that of single power-law fit. The index of high-energy end is $ \alpha_{T, high} = 2.69\pm0.24 $.

From Figures 2 and 3, we have found a similar power-law
dependence of the occurrence rate (energy and duration) for solar radio bursts, FRB 121102 and SGR 1935+2154. Interestingly,
solar X-ray flares in temporal with solar type III radio bursts also show a similar
power-law distribution \citep{Crosby1993,Wang2013}.
These distributions can be well understood within the
statistical framework of a self-organized criticality (SOC)
system \citep{Katz1986,Bak1987,Lu1991}. The power-law
distributions of energy and duration naturally follow from such a model
since the system has no characteristic spatial
scale above the elementary scale of the smallest avalanche
(the smallest-energy event), up to the system size. The power-law indices
are consistent with the prediction of the fractal-diffusive SOC model in
3-dimension Euclidean space \citep{Aschwanden2012}. Interestingly, similar dimension
was also found in the pulses of prompt emission of gamma-ray bursts \citep{Lyu2020}.

What can we learn from the similar distributions between solar type
III radio bursts and FRBs? It is generally believed that type III
bursts arise from the nonlinear conversion of Langmuir waves at the
local plasma frequency by energetic electron beams accelerated
during solar
flares \citep{Ginzburg1958}.
Numerical simulations have revealed that solar radio bursts are
caused by particle acceleration episodes that result from bursty
magnetic reconnection. From observations, direct
evidences have been found that energetic electrons are accelerated
by magnetic reconnections, which also produce X-ray
flares \citep{Cairns2018}.
The similar energy and duration distributions support that both radio bursts and associated X-ray flares are triggered by magnetic reconnection
as speculated by some theoretical models \citep{Lyutikov2002,Katz2016,Lyutikov2020}.
The similar power-law distributions of solar radio bursts and hard X-ray flares, together with similar power-law distributions between
SGR X-ray flares and FRBs, strengthens the fact that X-ray and radio emissions are tightly correlated in these sources.

The third statistical property is the waiting time distribution,
which has been well studied in solar X-ray
flares \citep{Wheatland1998}, and X-ray flares in gamma-ray bursts \citep{Wang2013}.
The waiting time $\Delta t$ is defined as the time interval between
two successive bursts. This distribution provides extra constraints
on theoretical models. For example, avalanche models predict that
bursts occur independently \citep{Wheatland1998}.
We use Poisson
process to explain the waiting time distributions. For constant
burst rate, the waiting time follows the Poisson interval
distribution \citep{Wheatland1998}
\begin{equation}
\label{eq:cPdt} P(\Delta t) = \lambda e^{-\lambda \Delta t},
\end{equation}
where $\lambda$ is the burst rate. If the rate is time dependent,
the distribution can be treated as a piecewise constant Poisson
process consisting $N$ intervals with $\lambda_i$ and duration
$t_i$. The wait time distribution can be derived by \citep{Aschwanden2010}
\begin{equation}\label{eq:addPdt}
P(\Delta t) \simeq \frac{1}{\bar{\lambda}} \sum_{i=1}^N
\frac{t_i}{T} \lambda_i^2 e^{-\lambda_i \Delta t},
\end{equation}
where $\bar{\lambda}$ is the average burst rate and $T$ is the
duration of the observing period \citep{Wheatland1998}. Equation
(\ref{eq:addPdt}) can be transformed into a continues function
\begin{equation}
\label{eq:conPdt} P(\Delta t) = \frac{\int_0^T \lambda(t)^2
	e^{-\lambda(t)\Delta t } dt}{\int_0^T \lambda(t) dt}.
\end{equation}
For the exponentially growing occurrence rate $f(\lambda)=\lambda^{-1}\exp(-\lambda/\lambda_0)$ \citep{Aschwanden2010},
the waiting time distribution is
\begin{equation}
\label{eq:Pdt} P(\Delta t) = \frac{\lambda_0}{(1 + \lambda_0 \Delta	t)^2}.
\end{equation}
For large waiting times ($\Delta t\gg 1/\lambda_0$), it gives the
power-law limit $P(\Delta t)\approx \Delta t^{-2}$.

Figure 4 shows the occurrence rates as a function of waiting times for solar
type III radio bursts at 4995 MHz (left panel), FRB 121102 (middle
panel) and SGR 1935+2154 (right panel), respectively.
The fitting results from MCMC method using equation (\ref{eq:Pdt}) are shown as
solid lines in Figure 4. The mean rates are
$ \lambda_0=1.10^{+0.06}_{-0.05} \times 10^{-5} ~(6s)^{-1}$, $1.37^{+0.34}_{-0.26} \times 10^{-5} \rm ~ms^{-1}$
and $22.78^{+2.41}_{-2.30} \rm ~day^{-1}$ for solar radio bursts, FRB 121102 and SGR 1935+2154, respectively.
The waiting times in these sources can be understood by the same exponentially growing occurrence model.
The best-fitting parameters are shown in Table 1.
The same waiting time distributions of SGR 1935+2154 and FRB 121102 also support the tight correlation between radio and X-ray
emissions. The difference of mean rate $\lambda_0$ indicates that the magnetar powered FRB 121102 is much more active
than SGR 1935-2154. The central magnetar of FRB 121102 is estimated to be very young, about a few decades \citep{Metzger2017,Cao2017,Zhao2020}. The age of SGR 1935+2154 is uncertain, from 3000 years \citep{Israel2016}, to 16000 years \citep{Zhou2020}. According to \citet{Beloborodov2016}, the magnetic activity timescale is about 20 years in the direct Urca case (high-mass neutron star) and 700 years in modified Urca (normal-mass neutron star) case. Therefore, the mean rate of FRB 121102 is higher than that of SGR 1935+2154.

\section{Summary}
Many similar phenomena occur in astrophysical systems with spatial and magnetic field scales different by many orders of magnitudes.
In this paper, based on the observational facts that the radio bursts in the Sun, XTE J1810-197 and FRB 200428 are all
accompanied by X-ray flare emissions, we find a universal correlation $\log L_X=(1.12\pm 0.02)\times \log L_R+10.22\pm 0.26$ between X-ray luminosity
and radio luminosity over twenty orders of magnitude in these sources. This correlation implies that the
energetic electrons producing the X-ray flares and those causing the
radio emissions have a common origin. It must be noted that  in
the G\"{u}del-Benz correlation,  the radio emissions are gyrosynchrotron radiation. However,
the radio emission is from coherent process in our correlation. In future, this correlation can be tested using
more FRBs and associated X-ray flares. For solar flares, the arrive time of radio peak is delayed by about a few seconds relative to
the hard X-ray flare peak \citep{Bastian1998}. For FRB 200428, the two X-ray peaks
occur within about $t_\delta \sim$3 ms of the corresponding radio burst peaks \citep{CHIME/FRB2020,Li2020}. Therefore,
high-time resolution observations are needed to clarify this point \citep{Mereghetti2020,Margalit2020,Wu2020}. The time delay is also important to
distinguish physical models \citep{Lyutikov2020,Metzger2019}.

We also found similar statistical properties, including power-law distributions of energy, duration and waiting time,
among solar radio bursts, FRB 121102 and SGR 1935-2154. The universal $L_x-L_R$ correlation is also supported by the same energy and waiting time
distributions of SGR 1935-2154 and FRB 121102. The radio bursts and X-ray flares in the Sun and magnetar
SGR 1935+2154 are both show similar power-law distributions. These distributions can be well understood within the
statistical framework of a self-organized criticality system.
The universal correlation and the statistical similarities, together with the fact that
solar radio bursts and X-ray flares are powered by magnetic energy, reveal that
Galactic and cosmological FRBs are powered by magnetic energy of magnetars, and can be treated as scaled-up solar radio bursts.
It supports that magnetic activities govern the fundamental
physical processes in all of these systems. It must be noted that magnetars differ from solar radio souces in their many orders of magnitude higher rotation rate,
which creates much different electro-dynamic and magneto-hydrodynamic effects \citep[e.g.,][]{Sturrock2012}, which should be extensively studied.

\section*{Acknowledgements}
We thank the anonymous referee for constructive and helpful comments. This work is supported by the National Natural Science Foundation of China
(grants U1831207 and 11833003), and the National Key Research and Development
Program of China (grant 2017YFA0402600).

\section*{Data Availability}
The data used in the paper is included in the published papers as shown in Section 2.

\bibliographystyle{mnras}
\bibliography{ms}


\begin{figure*}
	\centering
	\includegraphics[width=0.75\linewidth]{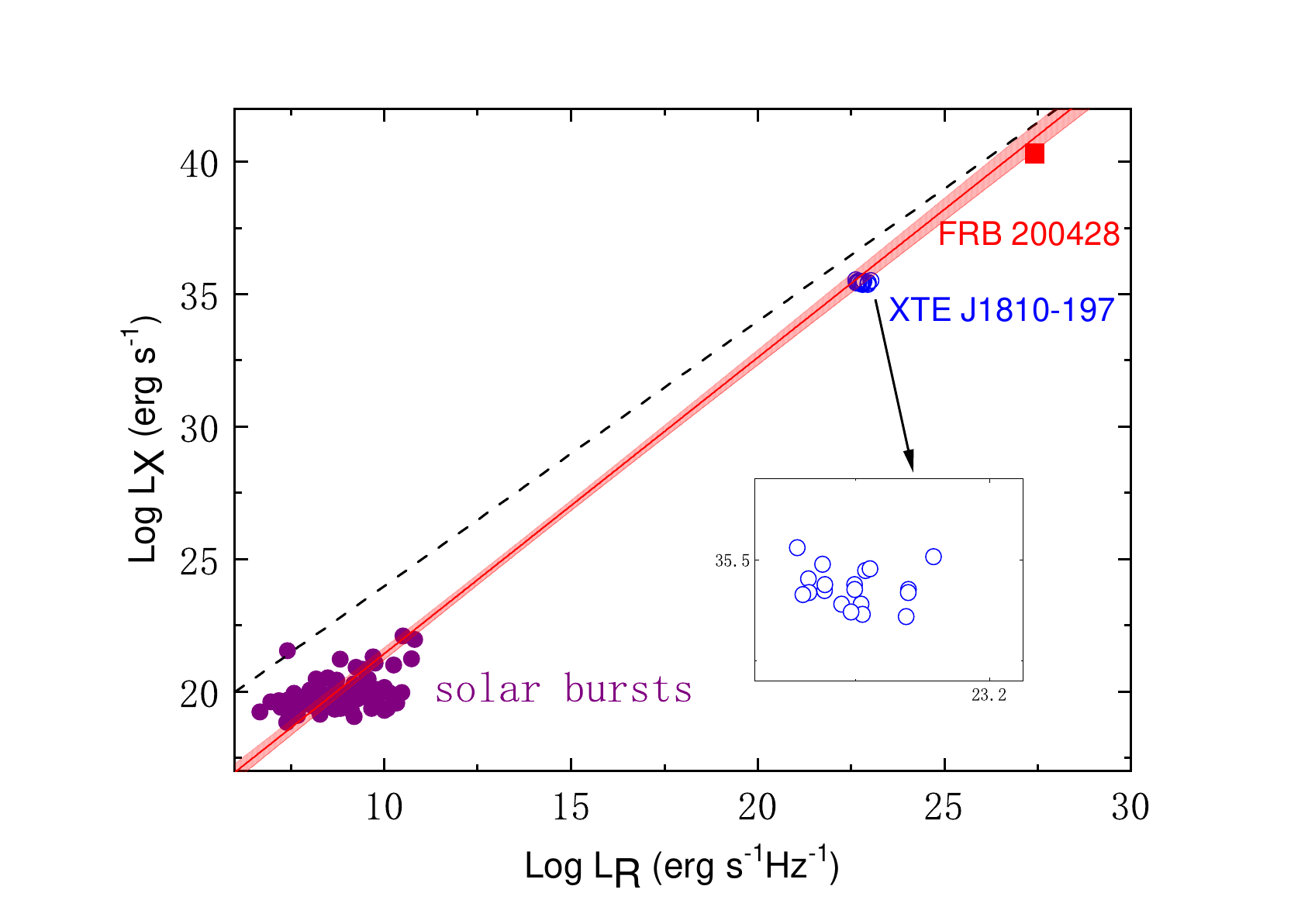}
	\caption{The $L_X-L_R$ correlation over twenty orders of magnitude. Purple points indicate solar type III radio bursts. X-ray and radio bursts from XTE J1810-197 are shown as blue open points. Red square shows the Galactic FRB 200428 and the associated X-ray flare. The red line is the fitting curve $\log L_X=(1.10\pm 0.01)\times \log L_R+(10.17\pm 0.10)$ (equation 1). FRB 200428 (red square) is dramatically consistent with this correlation. The dashed line is the G\"{u}del-Benz correlation, which only can be used for gyrosynchrotron radio emissions. Apparently, the coherent radio emissions deviate from the G\"{u}del-Benz correlation.}
	\label{LxLr}
\end{figure*}

\begin{figure*}
	\centering
	\includegraphics[width=\linewidth]{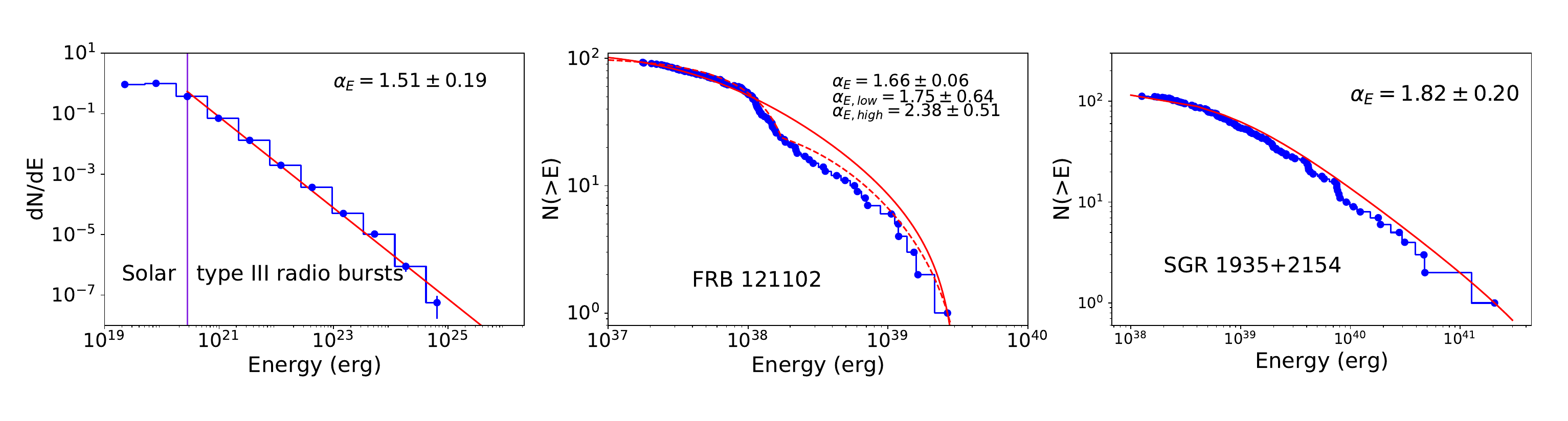}
	\caption{The occurrence frequency distribution of Energy. Error bars correspond
		to 68\% Poisson confidence intervals for the event numbers. Left panel: the differential distribution of energy for solar type III radio
		bursts is shown as step-wise blue curve. The data is observed by the
		United States Air Force Radio Solar Telescope Network (RSTN) from
		the National Centers for Environmental information (NCEI) between
		1979 and 2010. The best-fitting is shown as red line with power-law
		index $\alpha_E = 1.51\pm0.19$ ($1\sigma$). Middle panel: The step-wise blue
		curve represents the cumulative distribution of energy for FRB
		121102. We fit the cumulative distribution using a
		power-law function with a threshold $N(>E) = 1 + (N_{ev} - 1) \left( \frac{(E_2 + E_0)^{1 - \alpha_E} - (E + E_0) ^ {1 - \alpha_E}}{(E_2 + E_0)^{1 - \alpha_E} - (E_1 + E_0) ^ {1 - \alpha_E}} \right)$ with $ \alpha_E = 1.66\pm0.06$, which is shown
		as red line. We also show the fitting result of broken power-law in this panel with red dashed line. The power-law indices are also shown in this panel.
		 Right panel: The cumulative distribution of energy for SGR 1935+2154 with the best-fitting $\alpha_E = 1.82\pm0.20$.
		In order to avoid the selection effect, only the data above the
		break (vertical line) is fitted. The power-law indices are consistent with each other at $1\sigma$ confidence level.}
	\label{fig:frbE}
\end{figure*}

\begin{figure*}
	\centering
	\includegraphics[width=\linewidth]{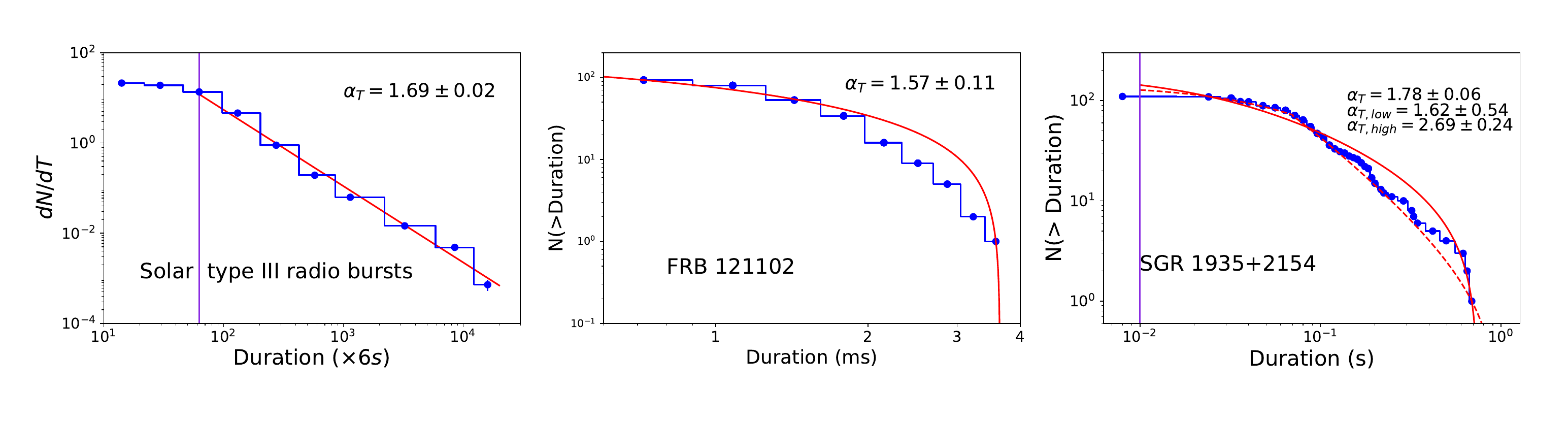}
	\caption{The frequency distribution of burst duration time. Left panel: we give the differential distribution of duration of solar
		type III radio bursts. The best fitting is
		$ \alpha_T = 1.69\pm0.02 $. Middle panel: The cumulative distribution of duration of FRB
		121102 is shown. The best fitting is $ \alpha_T = 1.57\pm 0.11$ (red line).
		Right panel: The cumulative distribution of duration of SGR 1935+2154 is shown with the best-fitting $\alpha_T = 1.78\pm 0.06$
		In order to avoid the selection effect, only the data above the
		break (vertical line) is fitted. The red dashed line is the
		fitting result of broken power-law model. We show the power-law indices in this panel.	
		Considering the $1\sigma$ uncertainties, the power-law indices are consistent.}
	\label{fig:frbwidth}
\end{figure*}

\begin{figure*}
	\centering
	\includegraphics[width=\linewidth]{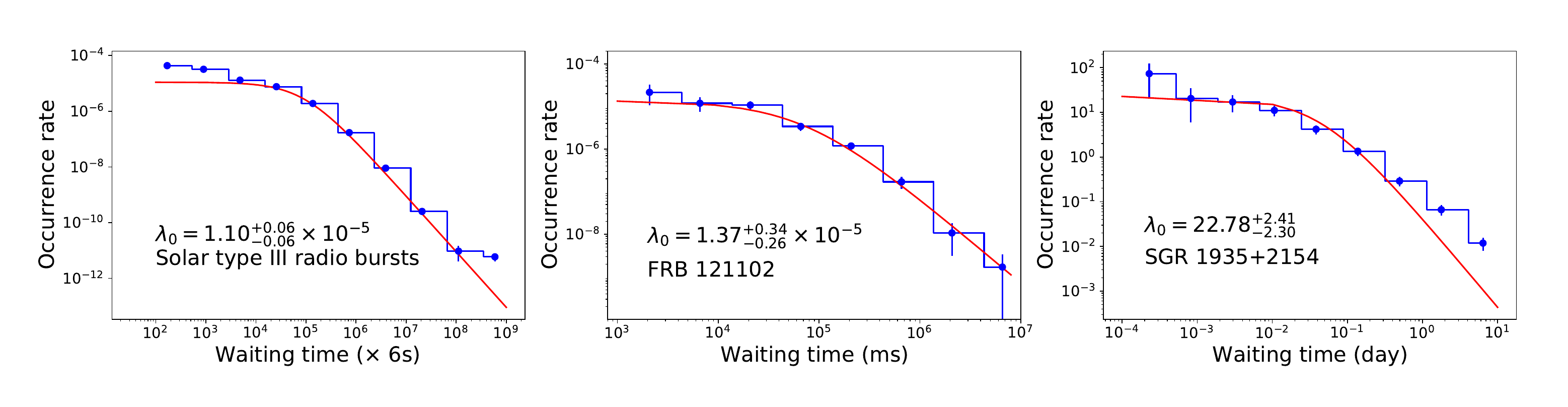}
	\caption{Waiting-time frequency distribution. Considering
		Poisson random process, we use $P(\Delta t) = \lambda_0 / (1 + \lambda_0 \Delta t)^2$ to fit the
		distributions. Left panel: The occurrence rate as a function of waiting time for solar type
		III radio bursts is shown as red line with best-fitting parameter $ \lambda_0=1.10^{+0.06}_{-0.05} \times 10^{-5}~ \rm (6s)^{-1}$.
		Middle panel: The occurrence rate as a function of waiting time for FRB 121102.
		The best-fitting parameter is $\lambda_0=1.37^{+0.34}_{-0.26}
		\times 10^{-5} \rm~ ms^{-1}$. Right panel: The occurrence rate as a function of waiting time for SGR 1935+2154.
		The best-fitting parameter is $\lambda_0=22.78^{+1.51}_{-1.46} \rm ~day^{-1}$. The waiting time distributions in these three sources
		can be explained by the same exponentially growing occurrence model.}
	\label{fig:frbwaiting}
\end{figure*}

\begin{table*}
	\centering \caption{The fitting results for
		solar radio bursts, FRB 121102 and SGR 1935+2154 with equation (3). The uncertainties are the 68.3\% confidence
		interval. \label{tab:1}}
	\begin{tabular}{|c|c|c|c|c|}
		\hline
		source & number of bursts  & $ \alpha_E $ 		& $\alpha_T$               &   $\lambda_0$ \\
		\hline
		solar radio bursts       &   2091		& $ 1.51 \pm 0.19 $	& $ 1.69 \pm 0.02 $ &   $1.10^{+0.06}_{-0.06} \times 10^{-5} ~(6s)^{-1}$   \\
		\hline
		FRB 121102             &   93            & $ 1.66 \pm 0.06 $	& $ 1.57 \pm 0.11 $ &   $ 1.37^{+0.34}_{-0.26} \times 10^{-5} \rm ~ms^{-1}$   \\
		\hline
		SGR 1935+2154      &   112            & $ 1.82 \pm 0.20 $	& $ 1.78 \pm 0.06 $  &   $ 22.78^{+2.41}_{-2.30} \rm ~day^{-1} $   \\
		\hline
	\end{tabular}
\end{table*}

\end{document}